\documentclass[12pt]{article}
\usepackage{amsmath}
\usepackage{graphicx}
\begin{document}

\title{Suppression of elliptic-flow-induced correlations in an observable of
possible local parity violation}
\author{Adam Bzdak\\
RIKEN BNL Research Center\\
Brookhaven National Laboratory\thanks{%
Address: Upton, NY 11973, USA; email: abzdak@bnl.gov}}
\date{}
\maketitle

\begin{abstract}
We show that fluctuations in elliptic anisotropy in heavy-ion
collisions can be used to significantly reduce the contribution of
transverse-momentum conservation, and of all background effects independent
on the orientation of the reaction plane, from an observable of the chiral
magnetic effect. We argue that for a given impact parameter, the magnetic
field is approximately independent of the fluctuating shape of the fireball.

\end{abstract}

\vskip0.6cm

\textbf{1.} The main feature of the chiral magnetic effect (CME) \cite%
{Kharzeev:2007jp} is the existence of an electric current parallel (or
antiparallel) to the direction of magnetic field produced in heavy-ion
collisions~\cite{Rafelski:1975rf}, see also \cite{Kharzeev:2007jp}, \cite%
{Skokov:2009qp}. Consequently, charge separation can be observed in the
direction perpendicular to the reaction plane. Recently, evidence for the
CME was found in lattice QCD calculations~\cite{Buividovich:2009my}. A
detailed discussion of this effect was given in Ref.~\cite{Kharzeev:2009fn}.

As discussed in Ref. \cite{Voloshin:2004vk}, such charge separation can be
observed via the following two-particle correlator%
\begin{equation}
\gamma =\left\langle \cos (\phi _{1}+\phi _{2}-2\Psi _{RP})\right\rangle ,
\label{gam-def}
\end{equation}%
where $\Psi _{RP},$ $\phi _{1},$ and $\phi _{2}$, respectively, denote the
azimuthal angles of the reaction plane, and the two charged particles
produced. Alternative observables were proposed in Refs.~\cite%
{Ajitanand:2010rc,Liao:2010nv} (see last section). Recently, the STAR
collaboration measured $\gamma $ and their result \cite{:2009uh,:2009txa} is
qualitatively consistent with the CME expectation; however, the
interpretation of these data still are debatable, see e.g., \cite%
{Pratt:2010zn,Schlichting:2010qia,Bzdak:2009fc,Bzdak:2010fd,Wang:2009kd,Muller:2010jd,Teaney:2010vd,Ma:2011um}%
. The interpretation of experimental results is not simple because
practically all two-particle correlations contribute to $\gamma $, as seen
from (here $\Psi _{RP}=0$) 
\begin{equation}
\gamma =\left\langle \cos (\phi _{1})\cos (\phi _{2})\right\rangle
-\left\langle \sin (\phi _{1})\sin (\phi _{2})\right\rangle .
\end{equation}%
Indeed, in the presence of elliptic anisotropy \cite{Voloshin:2008dg}, 
both terms can differ even though the correlation mechanism itself is 
not directly sensitive to the orientation of the reaction plane.

In this paper, we discuss this problem in detail. In the next Section, we
undertake explicit calculations to quantify the contribution of elliptic
anisotropy to $\gamma $. Next, we argue that in peripheral heavy-ion
collisions (where the CME is considered to have a maximum strength \cite%
{Kharzeev:2007jp}) we expect large fluctuations in elliptic flow ($v_{2}$)
such that it allows us to deduct the $v_{2}$-driven background from $\gamma $%
. We also show that the changing shape of the fireball, at a given impact
parameter, does not change the contribution of the CME to $\gamma $.
Finally, we present a Monte Carlo model, wherein we evaluate the
contribution of transverse-momentum conservation to $\gamma $ at vanishing
elliptic anisotropy ($v_{2}\rightarrow 0$). In the last section we give our
comments and conclusions.

\bigskip

\textbf{2.} By definition%
\begin{equation}
\gamma =\frac{\int \rho _{2}(\phi _{1},\phi _{2},x_{1},x_{2},\Psi _{RP})\cos
(\phi _{1}+\phi _{2}-2\Psi _{RP})d\phi _{1}d\phi _{2}dx_{1}dx_{2}}{\int \rho
_{2}(\phi _{1},\phi _{2},x_{1},x_{2},\Psi _{RP})d\phi _{1}d\phi
_{2}dx_{1}dx_{2}},  \label{gam-i}
\end{equation}%
where $\rho _{2}$ is the two-particle distribution at a given angle of the
reaction plane, $\Psi _{RP}$. To make our notation shorter, we denote $%
x=(p_{t},\eta )$ and $dx=p_{t}dp_{t}d\eta $, where $p_{t}$ is the absolute
value of transverse-momentum, while $\eta $ is pseudorapidity. The
distribution $\rho _{2}$ can be expressed via the correlation function $C$ 
\begin{equation}
\rho _{2}(\phi _{1},\phi _{2},x_{1},x_{2},\Psi _{RP})=\rho (\phi
_{1},x_{1},\Psi _{RP})\rho (\phi _{2},x_{2},\Psi _{RP})[1+C(\phi _{1},\phi
_{2},x_{1},x_{2})],  \label{ro2}
\end{equation}%
with the single-particle distribution\footnote{%
Higher $v_{n}$ results in terms proportional to $v_{2}v_{4}$, $v_{4}v_{6}$, 
etc., and can be neglected, see Eq. (\ref{gam-v2}).} 
\begin{equation}
\rho (\phi ,x,\Psi _{RP})=\frac{\rho _{0}(x)}{2\pi }[1+2v_{2}(x)\cos \left(
2\phi -2\Psi _{RP}\right) ],  \label{ro1}
\end{equation}%
where $\rho _{0}(x)$ does not depend on $\phi $ and $\Psi _{RP}$. We study
only those correlations that do not depend on the reaction plane, i.e., $C$
only depends on $\Delta \phi =\phi _{1}-\phi _{2}$. Next we expand $C$ in a
Fourier series%
\begin{equation}
C(\Delta \phi ,x_{1},x_{2})=\sum\nolimits_{n=0}^{\infty
}a_{n}(x_{1},x_{2})\cos \left( n\Delta \phi \right) ,  \label{c-fourier}
\end{equation}%
where $a_{n}(x_{1},x_{2})$ does not depend on $\phi _{1}$ and $\phi _{2}$.
Substituting (\ref{c-fourier}) and (\ref{ro2}) into Eq. (\ref{gam-i}), we
obtain%
\begin{equation}
\gamma =\frac{1}{2N^{2}}\int \rho _{0}(x_{1})\rho
_{0}(x_{2})a_{1}(x_{1},x_{2})[v_{2}(x_{1})+v_{2}(x_{2})]dx_{1}dx_{2},
\label{gam-v2}
\end{equation}%
where $N=\int \rho _{0}(x)dx$ and we assume that $1+a_{n}\approx 1$. As seen
from Eq. (\ref{gam-v2}), all correlations with non-zero $a_{1}(x_{1},x_{2})$
contribute to $\gamma $, even if the underlying correlation mechanisms do
not depend on the orientation of the reaction plane. This finding explains
why transverse-momentum conservation \cite%
{Pratt:2010zn,Bzdak:2010fd,Ma:2011um}, local charge-conservation \cite%
{Schlichting:2010qia}, resonance- (cluster-) decay \cite{Wang:2009kd}, and
all other correlations with $\Delta \phi $ dependence contribute to $\gamma $%
. However, as pointed out recently in Ref. \cite{Longacre:2011pd}, those
correlations can be removed from $\gamma $ by taking only those events where 
$v_{2}(x)\approx 0$. We note that $v_{2}(x)$ is defined solely through Eq. (%
\ref{ro1}), and it can be positive or negative.

Taking the CME into account\footnote{%
The value of $\chi $ flips between $-1$ and $+1$ so that $\frac{1}{2}%
\sum_{\chi }\rho _{\chi }=\rho $, defined in Eq. (\ref{ro1}).}%
\begin{equation}
\rho _{\chi }(\phi ,x,\Psi _{RP})=\frac{\rho _{0}(x)}{2\pi }\left[
1+2v_{2}(x)\cos \left( 2\phi -2\Psi _{RP}\right) +2\chi d(x)\sin \left( \phi
-\Psi _{RP}\right) \right] ,
\end{equation}%
we obtain,%
\begin{eqnarray}
\gamma &=&-\frac{1}{N^{2}}\left[ \int d(x)\rho _{0}(x)dx\right] ^{2}+  \notag
\\
&&\frac{1}{2N^{2}}\int \rho _{0}(x_{1})\rho
_{0}(x_{2})a_{1}(x_{1},x_{2})[v_{2}(x_{1})+v_{2}(x_{2})]dx_{1}dx_{2},
\label{gam-dv2}
\end{eqnarray}%
wherein the first term represents the CME, and the second term is the
elliptic-anisotropy-driven background.

\bigskip

\textbf{3.} We expected that elliptic anisotropy would be correlated with
the participant eccentricity $\epsilon _{2}$ \cite{Alver:2006wh}. 
In the center of mass of the
wounded nucleons, $\epsilon _{2}$ is given by 
\begin{equation}
\epsilon _{2}=\frac{\sqrt{\left( \sum_{i}r_{i}^{2}\cos (2\phi _{i})\right)
^{2}+\left( \sum_{i}r_{i}^{2}\sin (2\phi _{i})\right) ^{2}}}{%
\sum_{i}r_{i}^{2}},
\end{equation}%
wherein the wounded nucleons are characterized by their radii, $r_{i}$, and
their azimuthal angles, $\phi _{i}$.

\begin{figure}[h]
\begin{center}
\includegraphics[scale=0.45]{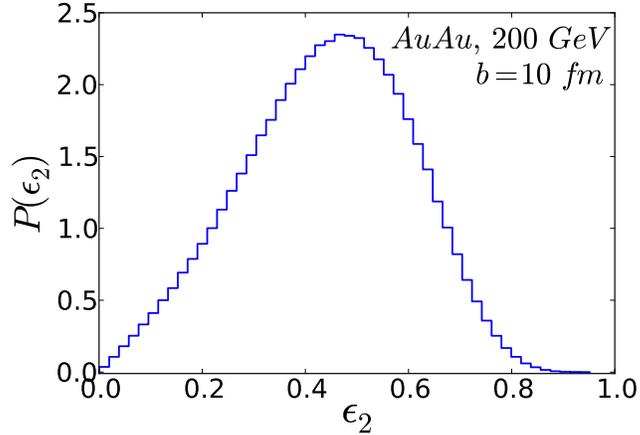}
\end{center}
\caption{Normalized density distribution of the participant eccentricity $%
\protect\epsilon _{2}$ at the impact parameter $b=10$ fm in Au+Au collisions
at $\protect\sqrt{s}=200$ GeV.}
\label{fig1}
\end{figure}

In Fig. \ref{fig1}, we present the calculated\footnote{%
We use the Monte Carlo Glauber calculation with standard parameters \cite%
{Alver:2008aq}.} $\epsilon _{2}$ distribution in Au+Au collisions at $\sqrt{s%
}=200$ GeV with $b=10$ fm (the impact parameter). Accordingly, even at $b=10$
fm, we obtain a broad range of $\epsilon _{2}$ (and $v_{2}$)\footnote{%
In $3$D event-by-event hydrodynamics \cite{BS}, $\left\langle
v_{2}\right\rangle =0.08$ and $[\left\langle v_{2}^{2}\right\rangle
-\left\langle v_{2}\right\rangle ^{2}]^{1/2}=0.04$ for $40-50\%$ centrality (%
$b\approx 10$ fm) in Au+Au collisions.} that can be used to significantly
change the background present in the second term of Eq. (\ref{gam-dv2}). 

\begin{figure}[h]
\begin{center}
\includegraphics[scale=0.45]{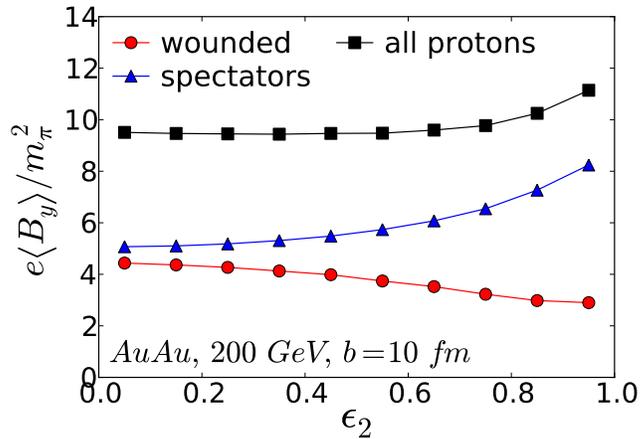}
\end{center}
\caption{The out-of-plane component of magnetic field at $b=10$ fm in Au+Au
collisions at $\protect\sqrt{s}=200$ GeV as a function of the participant
eccentricity $\protect\epsilon _{2}$. Contributions from the wounded- and
spectator- protons are depicted separately.}
\label{fig2}
\end{figure}

However, we wanted to remove this background under the condition that the
contribution from the CME is approximately unchanged. To verify this, we
calculated\footnote{%
The details of our calculation are presented in Ref. \cite{Bzdak:2011yy}.}
the out-of-plane component of the magnetic field $B_{y}$ at $t=0$ (time) as
a function of $\epsilon _{2}$. As seen in Fig. \ref{fig2}, the magnetic
fields from wounded- and spectator- protons\footnote{%
Both sources are expected to contribute differently during the
evolution of the fireball \cite{Kharzeev:2007jp}.} are approximately constant
(in comparison to $v_{2}$ that scales linearly with $\epsilon _{2}$) in the
broad range of $\epsilon _{2}$. We conclude that fluctuating $v_{2}$ in
peripheral collisions will allow us to study the $v_{2}$ dependence of $%
\gamma $ at an approximately constant strength of the CME. 

In Figs. \ref{fig1} and \ref{fig2} we made our calculations at a given impact
parameter. In an actual experiment, we can select our peripheral collisions,
e.g., by the number of particles produced at midrapidity. We consider that
our conclusions also will hold in this situation.

\bigskip

\textbf{4.} As an example, we calculated $\gamma $ as a function of $v_{2}$
in a model with only transverse-momentum conservation and elliptic
anisotropy.\footnote{%
The calculations presented in this section are for illustrative purposes
only, and should not be compared with the STAR data.} As shown in \cite%
{Pratt:2010zn,Bzdak:2010fd,Ma:2011um}, this effect contributes significantly
to $\gamma $ and reasonably describes the $p_{t}$ and $\eta $ dependence.

We sampled $N_{all}=50$ particles with $p_{t}$ according to the thermal
distribution $p_{t}e^{-p_{t}/T}$ with $2T=0.45$ MeV, and $\phi $ according
to $1+2v_{2}(p_{t})\cos (2\phi )$. We took $v_{2}(p_{t})=0.14p_{t}$ for $%
p_{t}<2$ GeV and $v_{2}(p_{t})=0.28$ for $p_{t}>2$ GeV, so that we obtained
the integrated $v_{2}\approx 0.06$. Next, we imposed transverse-momentum
conservation\footnote{%
We accepted only those events where the total (summed over $N_{all}$
particles) $|p_{t,x}|<0.5$ GeV and $|p_{t,y}|<0.5$ GeV.} and calculated $%
\gamma $ as a function of selected $v_{2}=\frac{1}{N_{obs}}%
\sum_{i=1}^{N_{obs}}\cos (2\phi _{i})$, where $N_{obs}$ is the number of
observed particles (selected randomly from $N_{all}$). Owing to the
statistical fluctuations, we obtained a broad range of $v_{2}$ that allows
us to test Eq. (\ref{gam-v2}). 
\begin{figure}[h]
\begin{center}
\includegraphics[scale=0.45]{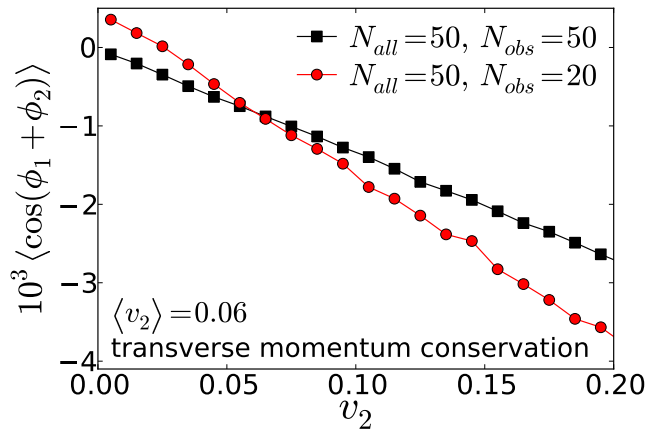}
\end{center}
\caption{The contribution of transverse-momentum conservation to $\protect%
\gamma $ as a function of $v_{2}$. The momentum is conserved for $N_{all}$
particles and $\protect\gamma $ is calculated for $N_{obs}$ (observed)
particles, selected randomly from $N_{all}$. }
\label{fig3}
\end{figure}

As seen from Fig. \ref{fig3}, $\gamma \rightarrow 0$ as $v_{2}\rightarrow 0$
for $N_{all}=N_{obs}=50$. However, it might be surprising that for $%
N_{obs}=20$, $\gamma $ becomes slightly positive as $v_{2}\rightarrow 0$. In
Ref. \cite{Bzdak:2010fd} the contribution of conserving transverse-momentum
was calculated analytically, and was shown to scale not only with $v_{2}$ in
the region where particles are observed, but also with $v_{2}$ in the whole
phase-space%
\begin{equation}
\gamma \sim -\frac{1}{N_{all}}\,\left[ 2\bar{v}_{2,\Omega }-\bar{\bar{v}}%
_{2,F}\right] \,,  \label{gam-an}
\end{equation}%
where $\bar{v}_{2}\sim \int \rho _{0}(x)v_{2}(x)p_{t}dx$, $\bar{\bar{v}}%
_{2}\sim \int \rho _{0}(x)v_{2}(x)p_{t}^{2}dx$ and $x=(p_{t},\eta )$, $%
dx=p_{t}dp_{t}d\eta $. The indexes $F$ and $\Omega $ indicate that
integrations are performed over full phase-space ($F$) or the phase space
wherein particles are measured ($\Omega $), respectively. When we calculate $%
\gamma $ for all produced particles, $\Omega =F$, then $v_{2,\Omega }=0$
implies $v_{2,\Omega }(p_{t},\eta )=0$ and $\bar{v}_{2,\Omega }=$ $\bar{\bar{%
v}}_{2,F}=0$. Consequently, $\gamma =0$ at $v_{2,\Omega }=0$. However, if we
measure only a fraction of all particles, then for $v_{2,\Omega }=0$ (and $%
\bar{v}_{2,\Omega }=0$), $\bar{\bar{v}}_{2,F}$ can differ from zero
(positive) and $\gamma \sim \bar{\bar{v}}_{2,F}/N_{all}>0$, as seen from Eq.
(\ref{gam-an}).

\bigskip

\textbf{5.} Several comments are warranted:

(i) Very recently, an observable related to $\gamma $ was studied as a
function of elliptic anisotropy \cite{Longacre:2011pd}. This finding argued
that for mid-peripheral Au+Au collisions $\gamma <0$ at $v_{2}=0$ for
same-charge pairs, which is consistent with the CME.

(ii) Even if $\gamma <0$ at $v_{2}=0$ for same-charge pairs, it does not
imply the existence of the CME. It only indicates the presence of some
correlation mechanism that explicitly depends on the orientation of the
reaction plane. To \textit{measure} the CME, a different observable is
needed, e.g., the multiparticle charge-sensitive correlator~\cite%
{Ajitanand:2010rc} or direct measurements of the electric dipole~\cite%
{Liao:2010nv}.

(iii) As argued in Ref. \cite{Voloshin:2010ut}, central U+U collisions also
can be used to distinguish between effects driven by elliptic anisotropy and
the CME. In the present paper, we considered only peripheral collisions,
where the CME and fluctuations in $v_{2}$ are expected to be the most
visible. Both methods can be used independently to reduce the contribution
of elliptic anisotropy.

(iv) In this paper, we proposed a way to remove the elliptic-flow-induced
background from the correlator $\gamma $. However, we note that in Ref.~\cite%
{Ajitanand:2010rc} a new multiparticle charge-sensitive correlator $C_{c}$
was proposed that is insensitive to correlations due to (elliptic) flow,
jets, or momentum conservation. The measurement of $C_{c}$ together with $%
\gamma $ (with removed background) could provide important information
about a possible signal of local parity violation.

In summary, we demonstrated that fluctuations in elliptic anisotropy in
peripheral heavy-ion collisions can be used effectively to reduce the
contribution of $v_{2}$-induced correlations from the two-particle
correlator (\ref{gam-def}). We showed that at a given impact parameter, the
magnetic field produced in heavy-ion collisions depends weakly on the
participant eccentricity, in contrast to the value of $v_{2}$. We also
discussed the contribution of transverse-momentum conservation to $\gamma $
at a vanishing $v_{2}$. Preliminary experimental analysis \cite%
{Longacre:2011pd} suggests the presence of a correlation mechanism that does
not scale with elliptic anisotropy.

\bigskip

\textbf{Acknowledgments}

We are grateful to Vladimir Skokov for numerous discussions. Correspondence
with Scott Pratt and Sergei Voloshin is highly appreciated. We thank Ron
Longacre for discussions about his recent paper. This investigation was
supported by the U.S. Department of Energy under Contract No.
DE-AC02-98CH10886 and by the grant N N202 125437 of the Polish Ministry of
Science and Higher Education (2009-2012).

\end{document}